\documentclass[usenatbib,usegraphicx, iop]{emulateapj}
\usepackage{amsmath}
\usepackage[]{hyperref}
\usepackage{graphicx}
\usepackage{color}
\usepackage{units}
\usepackage{ulem}
\usepackage{fixltx2e}
\DeclareGraphicsExtensions{.eps,.pdf,.png,.jpg}
\hypersetup{
    colorlinks=true,
    linkcolor=black,
    citecolor=black,
    filecolor=black,
    urlcolor=black,
}

\shorttitle{Sungrazing interstellar objects}
\shortauthors{Forbes and Loeb}

\begin{document}

\title{{\bf Turning up the heat on `Oumuamua}}

\author{John C. Forbes and Abraham Loeb}
\affil{Astronomy Department, Harvard University, 60 Garden St., Cambridge MA 02138, USA; \\ john.forbes@cfa.harvard.edu, aloeb@cfa.harvard.edu }

\begin{abstract}
We explore what may be learned by close encounters between extrasolar minor bodies like `Oumuamua and the Sun. These encounters may yield strong constraints on the bulk composition and possible origin of `Oumuamua-like objects. We find that such objects collide with the Sun once every 30 years, while about 2 pass within the orbit of Mercury each year. We identify preferred orientations for the orbits of extrasolar objects and point out known Solar System bodies with these orientations. We conclude using a simple Bayesian analysis that about one of these objects is extrasolar in origin, even if we cannot tell which.
\end{abstract}


\section{Introduction}

The detection of the interstellar object `Oumuamua on October 19, 2017 \citep{meech_brief_2017} was a surprise. Previous studies predicting the number density of interstellar comets in the galaxy were pessimistic that they would be detected even in the next generation of transient surveys \citep{moro-martin_will_2009}. The detection of even a single object like `Oumuamua immediately implied a vast abundance of interstellar objects \citep{laughlin_consequences_2017, do_interstellar_2018}, though their exact number density and size distribution remain uncertain.

Since its discovery, the physical nature of `Oumuamua has been debated, with different lines of evidence pointing to an asteroidal composition, while others point to a cometary origin. A priori `Oumuamua was expected to be an interstellar comet because solar systems likely have many icy bodies loosely bound in their outskirts. However, this picture was called into question when `Oumuamua showed no direct sign of cometary activity while observed in the solar system: no tails, and no dust \citep{micheli_non-gravitational_2018} or carbon-based molecules \citep{trilling_spitzer_2018}. These facts pointed away from a cometary origin, but are far from definitive. \citet{seligman_feasibility_2018} have argued that a comparatively thin layer of material could have insulated `Oumuamua's interior and prevented the traditional outgassing and brightening expected of comets. 

Recently \citet{micheli_non-gravitational_2018} discovered that the observed positions of `Oumuamua on the sky as it departs the solar system were highly significantly inconsistent with a purely hyperbolic orbit expected of purely gravitational forces. Non-gravitational accelerations are not uncommon among solar system comets \citep{rafikov_non-gravitational_2018}, and in fact the level of non-gravitational acceleration is within the range of values observed in the solar system. The favored explanation presented in \citet{micheli_non-gravitational_2018} is that there is directed outgassing, i.e. a jet, causing the non-gravitational acceleration, but because of unusual dust properties and carbon abundance, no dust \citep{micheli_non-gravitational_2018} or gas \citep{trilling_spitzer_2018} from `Oumuamua has been observed. \citet{rafikov_spin_2018} argues that a cometary jet of this sort should generically cause a rapid evolution in the spin of `Oumuamua that was not observed, unless there is extreme fine tuning of the lever arm of the jet's torque. Another possibility is that the standard estimates for `Oumuamua's albedo (both its magnitude and isotropy over the surface) are mistaken, and the object is not as elongated as naively inferred from the lightcurve. Another related possibility is that `Oumuamua is sufficiently reflective and low in mass, consistent with its lack of detection in Spitzer \citep{trilling_spitzer_2018}, that radiation pressure may affect its orbit \citep{bialy_could_2018, sekanina_preperihelion_2018}.

As the debate continues between asteroidal and cometary interpretations, it remains unclear if any firm conclusions can be reached as `Oumuamua itself has rapidly faded from view on its way out of the Solar System. \citet{hein_project_2017} has proposed a technically challenging mission that could use a gravitational assist from the sun to catch up to `Oumuamua for a flyby or in-situ measurement. More practically, \citet{seligman_feasibility_2018} have proposed preparing now for the likely future discovery of an interstellar object with a favorable orbit easily reachable with available rocketry. This would be particularly feasible for interstellar objects that were trapped in the Solar System by the gravitational ``fishing net'' of Jupiter and the Sun which may be identified through anomalous oxygen isotope ratios \citep{lingam_implications_2018} or their high inclination orbits \citep{siraj_identifying_2018}.

While in-situ exploration would no doubt settle the question of interstellar objects' compositions and yield other interesting discoveries, there may be a cheaper alternative. In this work we explore what may be learned by interstellar objects that happen to pass close to the Sun. In section \ref{sec:rate} we discuss the expected rate and orbital parameters of such objects, and in section \ref{sec:spectra} we comment on the compositional constraints that may be obtained in these events.

\section{Rates and Orbits}
\label{sec:rate}

The rate of encounters between the solar system and interstellar objects of number density $n$ moving with typical velocity $v$ is
\begin{equation}
\label{eq:basicrate}
\mathcal{R} = n \sigma v,
\end{equation}
where the cross-section $\sigma$ is dependent on both the velocity of the object far from the Sun and the maximum pericenter distance of interest $q_\mathrm{max}$,
\begin{equation}
\sigma = \pi q_\mathrm{max}^2 \left(1 + \frac{ 2 G M_\odot}{q_\mathrm{max} v^2} \right).
\end{equation}
Here the first term is the geometrical cross-section, and the second accounts for gravitational focusing.

\begin{figure}
\centering
\includegraphics[width= 9 cm]{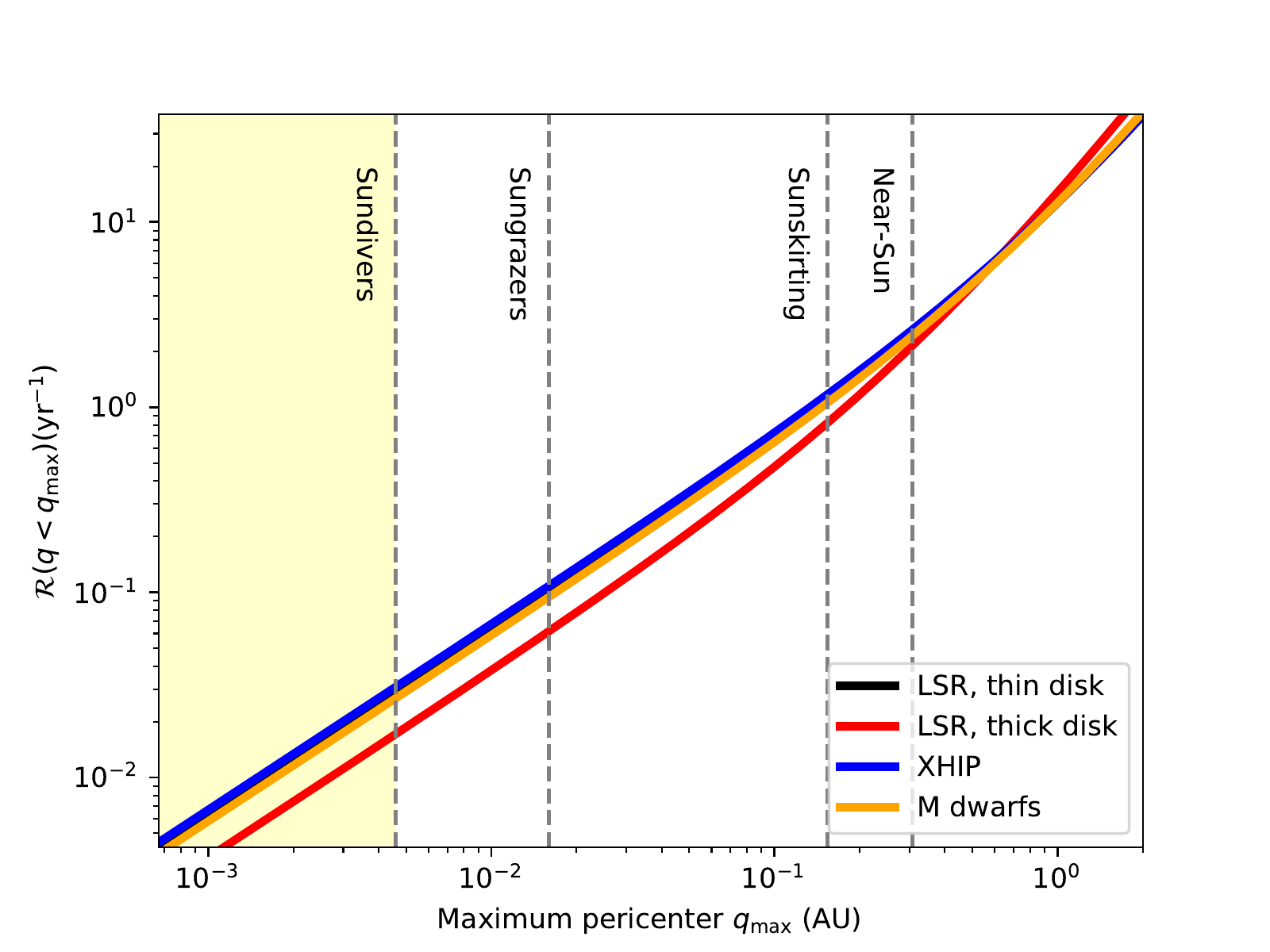}
\caption{Expected rates per year of interstellar objects with perihelion values less than $q_\mathrm{max}$. Each line shows a different assumption about the velocity distribution of interstellar objects - see Table \ref{tab:velocities} and text for details. Vertical lines indicate different categories of near-Sun objects, with objects that directly impact the Sun (Sundivers) shaded yellow. }
\label{fig:rate}
\end{figure}

The rate from Equation \eqref{eq:basicrate} can be generalized to account for the full velocity distribution, which we refer to as $f(\vec{v})$. In this case
\begin{equation}
\label{eq:rate}
\mathcal{R}(q<q_\mathrm{max}) = \int \int \int f(\vec{v}) \sigma(|\vec{v}|, q_\mathrm{max}) n |\vec{v}| d^3\vec{v},
\end{equation}
where the integrations are carried out over all possible values of the velocity vector. The distribution function of interstellar objects is tied to the velocity distribution of stars in the solar neighborhood, since the large abundance of interstellar objects inferred by `Oumuamua's detection necessitate contributions from most stars \citep{do_interstellar_2018}. The velocity distribution of stars in the solar neighborhood is roughly Gaussian, though there is a correlation between stellar age and velocity dispersion \citep[e.g.][]{nordstrom_geneva-copenhagen_2004}. The vertical velocity dispersion is smaller than the in-plane motion, and the velocity distribution is not centered on zero, but related to the Solar System's motion with respect to nearby stars. As discussed by \citet{mamajek_kinematics_2017}, `Oumuamua's inferred velocity as it entered the Solar System was not far from the canonical velocity of the Local Standard of Rest (hereafter LSR), namely the inverse of the Solar motion in the frame of the LSR, U,V,W = -7, -11, -10 km/s \citep{bland-hawthorn_galaxy_2016}. Its inferred velocity was even closer to the mean velocity of the \citet{reid_palomar/msu_2002} M dwarf sample and the XHIP sample of nearby stars \citep{anderson_xhip_2012}. These velocity distributions are summarized in Table \ref{tab:velocities}, though we note that the velocity distribution of nearby stars contains substantial substructure \citep[e.g.][]{trick_galactic_2018}, so this Gaussian model of $f(\vec{v})$ is an approximation. Our fiducial estimates will assume the first line of the table, namely a velocity centered on the LSR with a dispersion appropriate for the thin disk at an age of $\sim 10$ Gyr \citep{bland-hawthorn_galaxy_2016}.

\begin{table}
\caption{Plausible Velocity Distributions (units of km s$^{-1}$)}
\label{tab:velocities}
\begin{tabular}{lllllll}
 & U & V & W & $\sigma_U$ & $\sigma_V$ & $\sigma_W$ \\
\hline
LSR, thin disk & -10 & -11 & -7 & 35 & 25 & 25 \\
LSR, thick disk & -10 & -11 & -7 & 50 & 50 & 50 \\
XHIP & -10.5 & -18 & -8.4 & 33 & 24 & 17 \\
M dwarfs & -9.7 & -22.4 & -8.9 & 37.9 & 26.1 & 20.5 \\
     \end{tabular}
\end{table}

\begin{figure*}
\centering
\includegraphics[width= 15 cm]{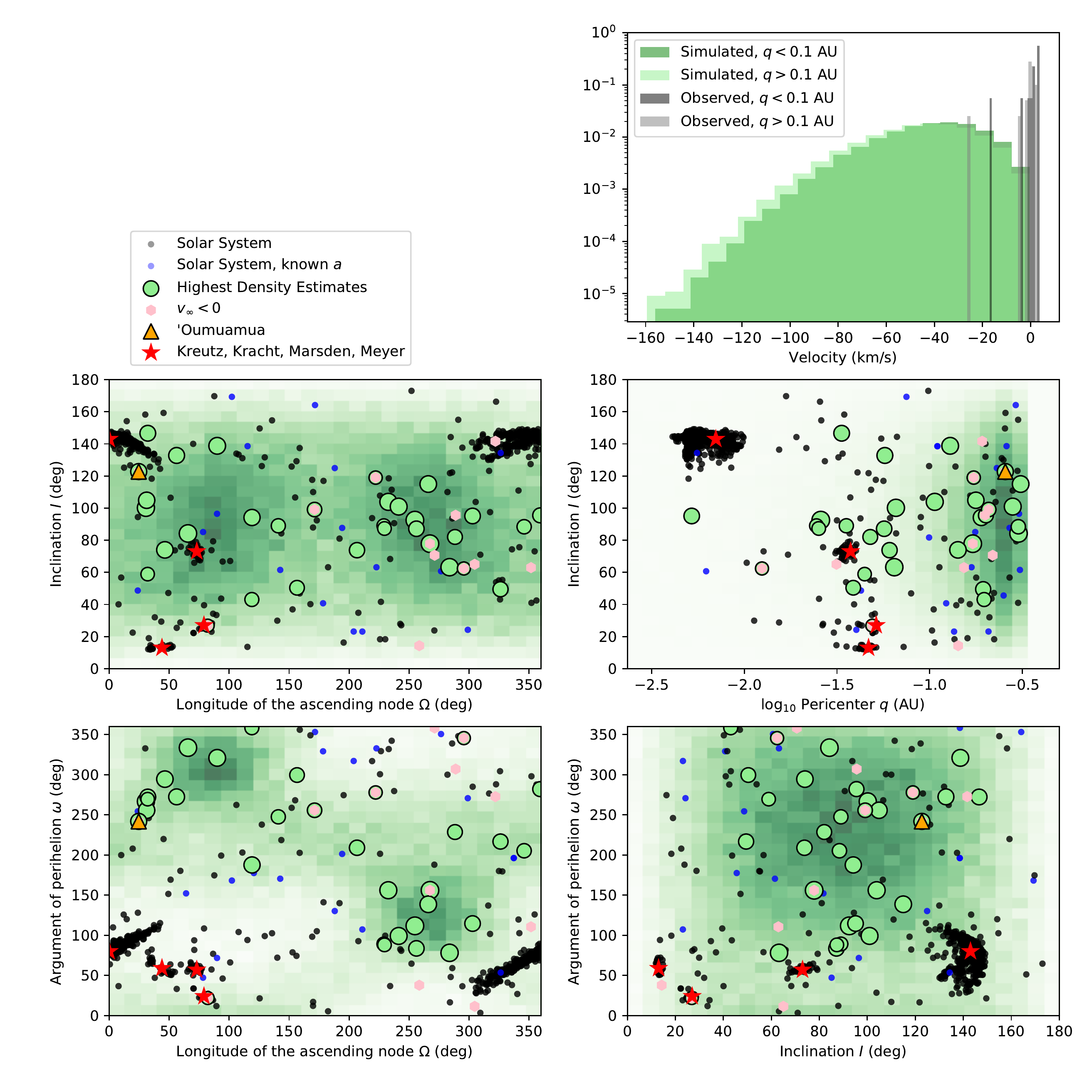}
\caption{Orbital Parameters. The bottom four panels show orbital elements for solar system bodies (small black or blue dots). Typical values for known Sungrazing or Sunskirting groups are marked as red stars, and `Oumuamua is shown as an orange triangle. The green histogram shows the distribution of simulated interstellar comets assuming a velocity distribution far from the Sun centered on the LSR velocity with velocity dispersions appropriate for the thin disk of the Milky Way. Green circles are Solar System objects that lie within the greatest density points as estimated by the 3D histogram in $\Omega, \omega,$ and $I$. The upper right panel shows the distributions of velocities $v_\infty$, with negative values meaning objects with hyperbolic orbits. Objects with $v_\infty<-1$ km/s are highlighted in pink in the lower four panels.}
\label{fig:orientations}
\end{figure*}

Figure \ref{fig:rate} shows the cumulative rate of encounters between interstellar objects and the Sun as a function of pericenter distance, with vertical dashed lines indicating various classes of Sungrazing comets as suggested in the review by \citet{jones_science_2018}. The boundaries in pericenter correspond to objects that hit the Sun (Sundivers), pass within the fluid tidal radius (Sungrazers), pass within half the semi-major axis of Mercury (Sunskirters), and pass within Mercury's orbit (near-Sun). The different colored lines show rates assuming different plausible velocity distributions, $f(\vec{v})$. While there is some difference, the uncertainty is dominated by the estimate of the interstellar density, which is based on the detection of a single object.

Assuming the \citet{do_interstellar_2018} estimate for the density of interstellar objects leads to rates of close encounters with the Sun between a few per year and about 1 per 30 years, depending on the exact pericenter distance. This raises a few interesting prospects, namely {\it the possibility that one or more near-Sun comets detected in the past few decades was interstellar in origin}, and {\it the high likelihood that many interstellar objects will have observable close encounters with the Sun in the coming years.}

To address these possibilities, we set out to determine the expected distribution of the orbits of interstellar objects. We do so via a Monte Carlo method. In particular, for a given value of $q_\mathrm{max}$, we first draw $N$ values of the velocity vector from its distribution $f(\vec{v}_i)$ where $i=1,...N$. Each of these samples is weighted by $w_i = \sigma(|\vec{v}_i |, q_\mathrm{max}) |\vec{v}_i| $. Then $M$ samples are drawn from this weighted distribution, i.e. for each draw (which we shall index by $j$) a given $\vec{v}_i$ is chosen with probability $w_i/\sum_{i=1}^N w_i$. We use $N=2\times10^5$ and $M=1\times 10^5$ throughout this work. Having specified the object's velocity far from the solar system $\vec{v}_j$, we now specify a spatial position to fully determine its orbit. %We will explicitly specify the object's position in right-handed galactic cartesian coordinates centered on the sun, where $\hat{x}$, $\hat{y}$ and $\hat{z}$ point parallel to the galactic plane towards the galactic center, parallel to the galactic plane the direction of the local circular velocity, and perpendicular to the galactic plane respectively. 

First, we determine the object's location in cylindrical coordinates with the positive axis of symmetry pointed in the direction of -$\vec{v}_j$. The object's location along this axis is taken to be an arbitrarily large number ($10^5$ AU). Within the plane specified by this `height' above the sun, a coordinate is drawn at random from a disk extending out to a maximum impact parameter 
\begin{equation}
b_\mathrm{max} = q_\mathrm{max} \sqrt{1+2 G M_\odot/(q_\mathrm{max} v^2)},
\end{equation}
i.e. $\sqrt{\sigma/\pi}$. This entails drawing a value of the cylindrical coordinate $\theta$ uniformly from 0 to $2\pi$, and the impact parameter is set by $b = b_\mathrm{max} \sqrt{u}$, where $u$ is drawn from a uniform distribution between 0 and 1. These coordinates are then transformed into galactic Cartesian coordinates, and finally both the position and velocity vectors are transformed from galactic to solar system cartesian coordinates via the rotation matrix defined in the Gaia DR1 documentation.\footnote{\url{https://gea.esac.esa.int/archive/documentation/GDR1/pdf/GaiaDR1_documentation_D.0.pdf} subsection 3.1.7.1.1}. Once transformed into the solar system Cartesian coordinates, i.e. the International Celestial Reference System (ICRS), we compute the standard orbital elements of these orbits, namely the semi-major axis, eccentricity, inclination, longitude of the ascending node, and argument of perihelion $a$, $e$, $i$, $\Omega$, and $\omega$, respectively. 

Figure \ref{fig:orientations} shows the distribution of these simulated orbits in the green histogram. For comparison we show the sample of comets and asteroids from the JPL small-body database\footnote{\url{https://ssd.jpl.nasa.gov/sbdb_query.cgi}} with pericenter $q<0.3$ AU (to focus on near-sun objects), and with orbits determined to be parabolic or hyperbolic\footnote{Queries were limited to be members of Parabolic Asteroid, Hyperbolic Asteroid, Asteroid (other), Hyperbolic Comet, Parabolic Comet, or Comet (other) }. 

The vast majority of known small Solar System bodies in this regime belong to the Kreutz group of Sungrazing comets \citep{kreutz_untersuchungen_1888}. This and other groups are clearly visible as clusters in Figure \ref{fig:orientations}, with orbital parameters of each from \citet{jones_science_2018} marked as red stars. These groups are each likely the result of a large progenitor body that has since fragmented. These groups are well-separated from the expected distribution of interstellar objects, though this is sensitive to assumptions about $f(\vec{v})$. `Oumuamua itself (shown as the orange triangle) is not in the peak of the distribution, but is within the expected range. To quantify this, we estimate the probability density over $\Omega$, $\omega$, and $I$ by filling a histogram with the $M$ sample orbits (the same histogram shown in Figure \ref{fig:orientations}). The resulting probability distribution $f(\Omega,\omega,I)$ is normalized so that  
\begin{equation}
\int_0^{2 \pi} \int_0^{2\pi} \int_0^{\pi} f(\Omega,\omega,I) d\Omega d\omega dI = 1,
\end{equation}
We highlight known small solar system bodies whose orbital elements give them values of $f(\Omega,\omega,I)$ greater than two thirds of `Oumuamua's value of $f$, as light green circles in Figure \ref{fig:orientations}, and in Table \ref{tab:interesting}. In addition to comets with large values of $ f(\Omega, \omega, I)$, we also include two near-Sun comets with $v_\infty<-1$ km/s, where $v_\infty = \mathrm{sign}(a) \sqrt{GM_\odot/|a|}$. 

We also estimate the probability that any one is extrasolar based purely on the orientation of its orbit (obviously the energy of the orbit would be a stronger indicator, but this is more difficult to measure). We apply Bayes' Theorem, 
\begin{equation}
P( ES | \vec{\phi}_i) = \frac{f(\vec{\phi}_i | ES) P(ES)}{f(\vec{\phi}_i | ES) P(ES) + f_\mathrm{ISO}(\vec{\phi}_i  | \mathrm{LPC}) P(\mathrm{LPC})},
\end{equation}
where the angles have been abbreviated $\vec{\phi}_i = (\Omega_i, \omega_i, I_i)$, $P(ES)$ is the prior probability that a given object is extrasolar, $P(\mathrm{LPC})$ is the prior probability that a given object is a long-period comet (hereafter LPC), and $f_\mathrm{ISO}$ is constructed in the same way as $f(\vec{\phi})$, except that the velocity distribution from which the simulated comets are drawn is centered at zero and isotropic with $\sigma_U=\sigma_V=\sigma_W=1$ km/s.

\begin{table*}
\caption{Known comets that line up well with the expected properties of interstellar objects}
\label{tab:interesting}
\begin{tabular}{lllllllll}
Full Name & q (AU) & e & I (deg) & $\Omega$ (deg) & $\omega$ (deg) & v$_\infty$ (km/s)\footnote{Based purely on instantaneous estimates of $a$; comets with $v_\infty<0$ may in fact be bound to the Solar System.}  & $4\pi^3 f(\vec{\phi})$ & $p(ES | \vec{\phi})$\footnote{Assuming a prior $P(ES)=0.01$} \\
\hline
     C/1865 B1 (Great southern comet) & 0.03 & 1.0 & 92.49 & 254.83 & 111.72 & - & 4.17 & 0.0338 \\ 
     C/2006 P1 (McNaught) & 0.17 & 1.0000190685 & 77.84 & 267.41 & 155.97 & -0.31 & 3.43 & 0.026 \\ 
     C/2011 L4 (PANSTARRS) & 0.3 & 1.00003268452 & 84.21 & 65.67 & 333.65 & -0.31 & 3.06 & 0.0192 \\ 
     C/1689 X1 & 0.06 & 1.0 & 63.2 & 283.75 & 78.13 & - & 2.87 & 0.0181 \\ 
     C/1970 B1 (Daido-Fujikawa) & 0.07 & 1.0 & 100.18 & 30.61 & 266.65 & - & 2.69 & 0.0131 \\ 
     C/1665 F1 & 0.11 & 1.0 & 103.89 & 232.7 & 156.09 & - & 2.69 & 0.0205 \\ 
     C/1677 H1 & 0.28 & 1.0 & 100.93 & 241.32 & 99.16 & - & 2.69 & 0.0238 \\ 
     C/1910 A1 (Great January comet) & 0.13 & 0.999995 & 138.78 & 90.04 & 320.91 & 0.19 & 2.59 & 0.0535 \\ 
     C/1884 A1 (Ross) & 0.31 & 1.0 & 114.98 & 265.99 & 138.6 & - & 2.5 & 0.0222 \\ 
     C/1851 U1 (Brorsen) & 0.14 & 1.0 & 73.99 & 46.43 & 294.45 & - & 2.5 & 0.0265 \\ 
     C/1577 V1 & 0.18 & 1.0 & 104.88 & 31.24 & 255.67 & - & 2.41 & 0.0198 \\ 
     C/2002 X5 (Kudo-Fujikawa) & 0.19 & 0.999842855646 & 94.15 & 119.07 & 187.58 & 0.86 & 2.32 & 0.0131 \\ 
       `Oumuamua (A/2017 U1) & 0.26 & 1.2011337961 & 122.74 & 24.6 & 241.81 & -26.41 & 2.22 & - \\ 
     C/1997 B4 (SOHO) & 0.06 & 1.0 & 132.78 & 56.2 & 272.4 & - & 2.04 & 0.0129 \\ 
     C/2000 J2 (SOHO) & 0.03 & 1.0 & 146.61 & 32.19 & 272.21 & - & 2.04 & 0.0241 \\ 
     C/2001 N1 (SOHO) & 0.01 & 1.0 & 95.09 & 302.92 & 114.59 & - & 2.04 & 0.0137 \\ 
     C/2000 Q1 (SOHO) & 0.06 & 1.0 & 87.14 & 256.11 & 83.66 & - & 1.94 & 0.0131 \\ 
     C/1882 F1 (Wells) & 0.06 & 0.999993648982 & 73.8 & 206.59 & 208.98 & 0.3 & 1.85 & 0.0125 \\ 
     C/1945 W1 (Friend-Peltier) & 0.19 & 1.0 & 49.48 & 326.2 & 216.71 & - & 1.85 & 0.0125 \\ 
     C/1859 G1 (Tempel) & 0.2 & 1.0 & 95.49 & 359.31 & 282.0 & - & 1.76 & 0.0126 \\ 
     C/2007 M8 (SOHO) & 0.05 & 1.0 & 82.04 & 288.26 & 228.58 & - & 1.76 & 0.0091 \\ 
     C/2005 Q6 (SOHO) & 0.04 & 1.0 & 50.4 & 156.59 & 299.54 & - & 1.76 & 0.0209 \\ 
     C/2017 S3 (PANSTARRS) & 0.21 & 1.00476984465 & 99.1 & 171.2 & 255.67 & -4.51 & 1.76 & 0.0086 \\ 
     C/1989 W1 (Aarseth-Brewington) & 0.3 & 1.00009317367 & 88.39 & 345.91 & 205.26 & -0.52 & 1.67 & 0.0128 \\ 
     C/2005 M3 (SOHO) & 0.04 & 1.0 & 88.99 & 140.98 & 247.5 & - & 1.67 & 0.01 \\ 
     C/1975 V1-A (West) & 0.2 & 0.999971 & 43.07 & 118.92 & 358.43 & 0.36 & 1.57 & 0.0187 \\ 
     C/2000 Y7 (SOHO) & 0.02 & 1.0 & 89.02 & 228.93 & 89.13 & - & 1.57 & 0.0094 \\ 
     C/2000 Y6 (SOHO) & 0.03 & 1.0 & 87.3 & 229.47 & 88.03 & - & 1.57 & 0.0094 \\ 
     C/1874 D1 (Winnecke) & 0.04 & 1.0 & 58.89 & 32.05 & 269.51 & - & 1.57 & 0.0187 \\ 
     C/1853 R1 (Bruhns) & 0.17 & 1.000664 & 119.0 & 222.12 & 277.84 & -1.85 & 1.3 & 0.0074 \\ 
     C/2012 S1 (ISON) & 0.01 & 1.00020100383 & 62.4 & 295.65 & 345.53 & -3.78 & 1.11 & 0.0052 \\ 
     \hline
     \end{tabular}
\end{table*}

To estimate $P(ES)$, we compare the rate of extrasolar objects (Equation \ref{eq:rate}) to the observed rate of dynamically new comets. For now we shall proceed under the naive assumption that the two size distributions are similar leaving $P(ES)$ independent of size. `Oumuamua has an absolute magnitude of $H_{10} \approx 22.08$ \citep[e.g.][]{bolin_APO_2018}. If the cumulative number density of comets is proportional to $10^{\alpha H_{10}}$ with $\alpha\approx 0.28$ \citep{hughes_cometary_1988, weissman_size_2001}, then extrapolating the flux of comets of $\sim 1\ \mathrm{yr}^{-1}$ within 5 AU for $H_{10} < 7$ to `Oumuamua-sized objects, we find a rate of about $1.7 \times 10^4\ \mathrm{yr}^{-1}$, though the rate may be closer to $10^{3}\ \mathrm{yr}^{-1}$ if the slope is shallower, with $\alpha = 0.2$ as per \citet{fernandez_magnitude_2012}. For different assumptions about $f(\vec{v})$ the expected rate of `Oumuamua-sized objects with pericenters $q<5 AU$ is about 200 yr$^{-1}$, keeping in mind the order of magnitude uncertainty from the background density being estimated from a single object. This implies that $P(ES)$ should be between 0.01 and 0.2.

Both interstellar objects and LPCs are subject to gravitational focusing, but the lower velocity of the LPCs enhances their rate by an additional factor of 2. Since LPCs also brighten as they approach pericenter, there may be a bias in composition. If extrasolar objects are rocky or have a layer of material protecting more volatile-rich interiors, they may be harder to discover than LPCs, in which case the range of $P(ES)$ estimated purely on the rates may be more like an upper limit. We therefore conservatively adopt $P(ES) \la 0.01$ as the plausible range for the prior, with lower values corresponding to rockier compositions, lower intrinsic rates, or steeper faint-end brightness distributions for LPCs. Table \ref{tab:interesting} includes estimates of the posterior probability that a given object is extra-solar in origin $p(ES|\vec{\phi})$ under the assumption that $P(ES)=0.01$ for all objects.

\section{Close Encounters with the Sun}
\label{sec:spectra}

There is a long history of the spectroscopic and narrow-band study of cometary tails with ground-based telescopes \citep{ahearn_ensemble_1995, fink_taxonomic_2009, langland-shula_comet_2011}.  Generally these studies are able to classify comets into different groups depending on the inferred production rates of H$_2$O, C$_2$, CN, and NH$_2$ as well as dynamical properties, which likely reflect formation in different parts of the protoplanetary disk \citep{levison_comet_1996}. Two comets, 96P/Machholz 1 \citep{schleicher_extremely_2008, langland-shula_unusual_2007} and Yanaka (1998r) have been shown to have highly depleted CN and C$_2$ relative to water, leading some to conjecture that they are interstellar in origin \citep{de_la_fuente_marcos_where_2018}. The promise of using close encounters with the sun to learn about extrasolar small bodies is that the sun has the ability to disrupt even large cometary nuclei via its intense radiation, sublimating not just surface volatiles but even silicates and iron. In principle this exposes the interiors of these objects to remote spectroscopy, which could place strong constraints on the composition of these objects. 

This has been done a few times, notably with the Atmospheric Imaging Assembly \citep[AIA][]{lemen_atmospheric_2012}, an EUV imager aboard the Solar Dynamics Observatory (SDO). AIA successfully observed comet C/2011 N3 (SOHO) in the process of being destroyed \citep{schrijver_destruction_2012}, and observations of comet C/2011 W3 Lovejoy were used as a probe of magnetic structure in the solar corona \citep{downs_probing_2013, raymond_solar_2014}. Lines of variously-ionized iron, oxygen, and carbon happen to lie in the AIA bands which are centered around high-ionization lines of iron common in the solar corona \citep[e.g.][]{bryans_extreme-ultraviolet_2012}. As the comet loses mass, the iron and oxygen in the tail are ionized by the corona and emit light in the different AIA bands. This process is nontrivial to model, and occurs over only about a minute. These measurements therefore yield only broad indications of the comet's composition. The observations tend to be well-fit by a model with standard cometary abundances \citep{bryans_extreme-ultraviolet_2012, pesnell_time-dependent_2014}. Observations by the Ultraviolet Coronograph Spectrometer (UVCS) \citep{kohl_ultraviolet_1995} aboard the Solar and Heliospheric Observatory (SOHO) have also been used to constrain the abundance of silicates sublimating from C/2011 W3 \citep{raymond_comet_2018}, though UVCS is no longer operational. 

Directed observations by the AIA for the well-studied comet C/2012 S1 (ISON) yielded no detections, which serves as a cautionary tale for the study of sungrazing comets with this method; likely the nucleus lost so much mass that by the time it reached the AIA field of view its surface area, and hence volatile production rate, had decreased below the detectable level \citep{bryans_absence_2016} - this size was only modestly smaller than `Oumuamua's. Another telescope with the potential to observe sungrazing comets is the forthcoming  Daniel K. Inoue Solar Telescope\footnote{\url{https://dkist.nso.edu/}} (DKIST). DKIST will observe the sun at high spatial and temporal resolution, and is equipped with multiple spectro-polarimeters. Although its field of view is comparatively small, detection of sungrazing comets sufficiently early will likely enable the telescope to be pointed away from its usual target, as the AIA did to observe C/2012 S1. DKIST's capabilities in this realm may be limited by its lack of a coronograph \citep{jones_science_2018}, but its unprecedented sensitivity and resolution may yield interesting discoveries.

These examples illustrate the value of obtaining spectral or narrow band imaging from Sungrazing comets in the UV, but also illustrate some of the difficulties of doing so. The relatively small number of observations is the result of the challenge of identifying these comets sufficiently far in advance. Even Kreutz group comets, for which the approximate rate and orbital parameters are reasonably well-known, have rarely been identified in ground-based surveys \citep{knight_photometric_2010, ye_where_2014}. Undoubtedly this will change with the Large Synoptic Surey Telescope \citep[LSST][]{schwamb_large_2018}, and we can expect the detections of interstellar objects and Sungrazing comets to increase dramatically. Direct impacts with the sun, while rare, may also yield substantial information as the comet is rapidly destroyed \citep{brown_mass_2011, brown_destruction_2015}.

\section{Conclusion}

We have examined the rate and orientation of orbits for extrasolar comets that have perihelion distances within the orbit of Mercury. While the first confirmed interstellar object, ``Oumuamua, has unclear composition and origin, future interstellar visitors have the potential to shed light on this mystery, especially if they happen to pass near the sun. Experience with past Sungrazers suggests that discovery from the ground is a crucial component to obtaining compositional information about the deep interior of the object's nucleus, so LSST will likely prove crucial in this regard. Among known comets that have passed close to the sun, none are paritcularly likely to have been extrasolar in origin, and hence our observations of these comets sheds little light on the question of `Oumuamua's composition. Nonetheless, we expect that perhaps one of these comets may in fact be interstellar in origin. Future spectroscopy of the gas evaporated from such comets could shed new light on the nurseries of the large-than-expected population of `Oumuamua-like objects.

\section*{Acknowledgements}

The authors would like to thank John Raymond for helpful conversations. We made use of NASA's ADS, JPL's small-body database, and the arXiv. JCF is supported by an ITC Fellowship, and this work was supported in part by a grant from the Breakthrough Prize Foundation.

\clearpage

\end{document}